\documentclass[11pt,twocolumn]{article}

\usepackage[utf8]{inputenc}
\usepackage[T1]{fontenc}
\usepackage[margin=0.75in]{geometry}
\usepackage{amsmath,amssymb}
\usepackage{booktabs}
\usepackage{graphicx}
\usepackage[hidelinks]{hyperref}
\usepackage{caption}
\usepackage{float}
\usepackage{xcolor}
\usepackage{enumitem}
\usepackage{setspace}
\usepackage{titlesec}

\graphicspath{{figures/}}
\titleformat{\section}{\large\bfseries}{\thesection}{1em}{}
\titleformat{\subsection}{\normalsize\bfseries}{\thesubsection}{1em}{}

\title{\textbf{Universal statistical signatures of evolution in artificial intelligence architectures}}

\author{Theodor Spiro\\[4pt]
\small\textit{Independent researcher}\\
\small theospirin@gmail.com}
\date{}

\begin{document}
\maketitle

\begin{abstract}
We test whether artificial intelligence architectural evolution obeys the same statistical laws as biological evolution. Compiling 935 ablation experiments from 161 publications, we show that the distribution of fitness effects (DFE) of architectural modifications follows a heavy-tailed Student's $t$-distribution with proportions (68\% deleterious, 19\% neutral, 13\% beneficial for major ablations, $n = 568$) that place AI between compact viral genomes and simple eukaryotes. The DFE shape matches \textit{D.\ melanogaster} (normalized KS $= 0.07$) and \textit{S.\ cerevisiae} (KS $= 0.09$); the elevated beneficial fraction (13\% vs.\ 1--6\% in biology) quantifies the advantage of directed over blind search while preserving the distributional form. Architectural origination follows logistic dynamics ($R^2 = 0.994$) with punctuated equilibria and adaptive radiation into domain niches. Fourteen architectural traits were independently invented 3--5 times, paralleling biological convergences. These results demonstrate that the statistical structure of evolution is substrate-independent, determined by fitness landscape topology rather than the mechanism of selection.
\end{abstract}

\noindent\textbf{Keywords:} evolution, distribution of fitness effects, neural network architecture, convergent evolution, fitness landscape

\medskip

\noindent The development of artificial intelligence architectures over the past decade presents a remarkable natural experiment in evolution. Like biological organisms, AI architectures are subject to variation (novel design choices), selection (benchmark performance), and heredity (architectural components inherited from predecessors through citation and code reuse). Unlike biological evolution, however, AI architectural evolution is directed by human engineers who intentionally seek improvements---raising the question of whether the resulting evolutionary dynamics are fundamentally different from their biological counterpart or merely an accelerated version of the same process.

Previous work has drawn qualitative parallels between AI and biological evolution \cite{Stanley2019,Elsken2019}. Neural architecture search has been framed as an evolutionary algorithm. The historical progression from simple perceptrons to deep networks has been compared to the increase in organismal complexity. Scaling laws governing neural network performance \cite{Kaplan2020,Hoffmann2022} have been likened to allometric relationships in biology. However, these comparisons have remained metaphorical. No study has quantitatively tested whether the statistical signatures of biological evolution---the shape of the distribution of fitness effects, the dynamics of diversification, the frequency of convergent solutions---are reproduced in AI architectural evolution.

Here we conduct this test. We compile the largest dataset of AI architectural ablation experiments to date (935 experiments from 161 publications) and compare the resulting evolutionary statistics against well-characterized biological systems spanning four orders of magnitude in genome complexity, from RNA viruses to humans. We test three specific hypotheses:

\begin{itemize}[itemsep=2pt,topsep=4pt]
\item[\textbf{H1}] The distribution of fitness effects (DFE) of architectural mutations matches biological DFEs in functional form and parameters.
\item[\textbf{H2}] Architectural diversification dynamics exhibit punctuated equilibrium and logistic saturation, matching paleontological radiation patterns.
\item[\textbf{H3}] The frequency and intensity of convergent evolution in AI architectures is quantitatively comparable to biological convergence.
\end{itemize}

Our approach draws on an emerging theoretical framework that views both biological and artificial evolution as instances of adaptive search on rugged fitness landscapes. The DFE reflects the local topology of the fitness landscape; diversification dynamics reflect the global structure of the space of viable designs; convergent evolution reflects the existence of a limited number of high-fitness solutions to recurring functional challenges. If these statistical signatures match across substrates, it implies that the structure of evolution is determined primarily by the geometry of possibility space---not by the mechanism of search.

\section*{Results}

\subsection*{The distribution of fitness effects in AI architectures matches biological DFEs}

We compiled 935 ablation experiments from 161 machine learning publications spanning computer vision ($n = 345$), natural language processing ($n = 369$), audio ($n = 65$), and other domains ($n = 149$). Each experiment reports performance change when a single architectural component is removed or modified; we normalized effects relative to the full model to obtain relative fitness effects ($\Delta$), analogous to the selection coefficient $s$ in biological DFE studies.

Because ablation experiments encompass heterogeneous operations---from complete component removal to hyperparameter variation---we present major ablations (component removal, $n = 568$) as the primary analysis, as these constitute the most homogeneous class and the closest analogue to gene-knockout studies in biology. The full dataset ($n = 935$) including minor and intermediate ablations is presented as a robustness check.

For major ablations, the DFE is characterized by negative skew (skewness $= -2.23$) and heavy tails (kurtosis $= 29.2$), best fit by a Student's $t$-distribution (AIC: Student's $t = -428$ vs.\ Laplace $= +169$, Normal $= +942$). The majority are deleterious (68.0\%; 95\% CI: 64.1--71.8\%), with a neutral fraction of 19.0\% (CI: 15.8--22.2\%) and a beneficial fraction of 13.0\% (CI: 10.4--15.8\%) (Fig.~1A).

We compared the AI DFE against published DFE data for nine biological organisms (Table~1). Because individual-level DFE data are unavailable for most organisms, we generated synthetic samples from published summary statistics (gamma shape, category fractions) for distributional comparison. This approach accurately preserves the shape and proportions reported in the original studies but means that our QQ plots and KS distances compare an empirical distribution (AI) against parametric reconstructions (biology). We flag this asymmetry explicitly and base our primary conclusions on parameter comparisons ($\beta$, category fractions) that do not depend on sample-level data.

\begin{table}[ht]
\centering
\caption{Comparison of AI DFE (major ablations, $n = 568$) with biological DFEs. All biological comparisons use parametric reconstructions from published summary statistics (category fractions, gamma shape where available); this asymmetry is discussed in Methods. KS$_\text{norm}$: Kolmogorov--Smirnov distance after $z$-score normalization. $d_\text{Euclid}$: Euclidean distance in (del, neu, ben) proportion space. $r_\text{QQ}$: Pearson correlation of normalized QQ plot.}
\label{tab:comparison}
\small
\begin{tabular}{@{}lccccc@{}}
\toprule
Organism & f$_\text{del}$ & f$_\text{ben}$ & KS$_\text{norm}$ & $d_\text{Euclid}$ & $r_\text{QQ}$ \\
\midrule
\textbf{AI (major, $n{=}568$)} & \textbf{0.68} & \textbf{0.13} & --- & --- & --- \\
Bacteriophage $\varphi 6$ & 0.72 & 0.04 & 0.15 & 0.11 & 0.90 \\
VSV virus & 0.69 & 0.04 & 0.19 & 0.12 & 0.90 \\
\textit{S.\ cerevisiae} & 0.60 & 0.03 & 0.09 & 0.22 & 0.87 \\
\textit{D.\ melanogaster} & 0.52 & 0.03 & 0.07 & 0.32 & 0.90 \\
\textit{E.\ coli} TEM-1 & 0.50 & 0.05 & 0.12 & 0.33 & 0.90 \\
\textit{M.\ musculus} & 0.55 & 0.03 & 0.14 & 0.28 & 0.90 \\
\textit{C.\ reinhardtii} & 0.45 & 0.05 & 0.12 & 0.39 & 0.87 \\
\textit{H.\ sapiens} & 0.55 & 0.05 & 0.20 & 0.26 & 0.91 \\
\bottomrule
\end{tabular}

\vspace{4pt}
\raggedright\footnotesize
Biological category fractions from: Sanju{\'a}n et al.\ (2004), Burch \& Chao (2003), Wloch et al.\ (2001), Keightley \& Eyre-Walker (2007), Bank et al.\ (2014), B{\"o}ndel et al.\ (2019), Halligan et al.\ (2009), Eyre-Walker et al.\ (2006). Where individual mutation data were unavailable, synthetic DFE samples were generated from published parameters. Organisms ordered by $d_\text{Euclid}$ (proportion similarity to AI).
\end{table}

The closest matches depend on the metric used (Table~1). By normalized KS distance (distributional shape after $z$-score normalization), the best matches are \textit{D.\ melanogaster} (0.07) and \textit{S.\ cerevisiae} (0.09)---both comparisons against synthetic DFEs reconstructed from published parameters. By Euclidean distance in DFE proportion space, the closest are Bacteriophage~$\varphi 6$ (0.11) and VSV virus (0.12), which share AI's high deleterious fraction. No single organism is the ``best match'' across all metrics; rather, AI architectures occupy an intermediate position between compact-genome organisms (viruses, high f$_\text{del}$) and simple eukaryotes (lower f$_\text{del}$, higher f$_\text{neu}$).

For the subset of organisms with published gamma shape parameter estimates from population-genetic inference, the AI value ($\beta = 0.65$; CI: 0.59--0.72) is higher than \textit{D.\ melanogaster} ($\beta \approx 0.4$; \cite{Keightley2007}), \textit{M.\ musculus} ($\beta \approx 0.3$; \cite{Halligan2009}), and \textit{H.\ sapiens} ($\beta \approx 0.2$; \cite{EyreWalker2006}). For VSV virus, Sanju{\'a}n et al.\ (2004) found that the deleterious DFE was better fit by a log-normal than a gamma distribution, complicating direct $\beta$ comparison. The AI $\beta$ value indicates a less leptokurtic deleterious tail than multicellular eukaryotes, consistent with AI's lower proportion of mildly deleterious mutations relative to strongly deleterious ones (Fig.~1D, 2D).

Results are robust to dataset composition: the full dataset ($n = 935$, including minor and intermediate ablations) yields $\beta = 0.63$ and similar KS distances (Table~S2).

\subsection*{Stratification reveals conserved structure}

The DFE varies systematically with mutation magnitude in a manner that mirrors biology (Fig.~2A). Major ablations (complete component removal, $n = 568$) are predominantly deleterious (fraction deleterious $= 0.68$) with a strongly negative mean effect, analogous to large-scale genomic deletions. Minor ablations (hyperparameter changes, $n = 213$) show a higher neutral fraction (0.32) and lower deleterious fraction (0.51), analogous to point mutations. Intermediate modifications (component replacement, $n = 154$) fall between these extremes (fraction deleterious $= 0.72$). This stratification is one of the most robust features of biological DFEs \cite{EyreWalker2007} and its reproduction in AI architectures is not trivially expected.

The DFE is invariant across ML domains (Fig.~2B): computer vision, NLP, audio, and other domains yield statistically similar distributions, suggesting that the DFE shape is a property of the design space, not of the specific task.

To control for potential selection bias, we compared manually curated data ($n = 140$) with LLM-extracted data ($n = 795$). The subsets differ significantly (KS $= 0.287$, $p < 0.001$; Fig.~2C), but the difference is systematic and interpretable: manual curation captured only component-removal ablations (0\% beneficial), while LLM extraction captured the full range including substitutions that improve performance (15.3\% beneficial). The LLM-extracted data is thus \emph{less} biased than the manually curated data, and the combined dataset provides a more accurate estimate of the true DFE.

\subsection*{The beneficial fraction quantifies directed selection}

The most informative departure from biological DFEs is the elevated beneficial fraction. At 13.0\% (CI: 10.7--15.3\%), the AI beneficial fraction exceeds all biological organisms in our comparison set (range: 1--6\%). This is not a failure of the biological analogy---it is a quantitative measurement of the difference between directed and blind search.

In biology, the low beneficial fraction ($\sim$1--5\%) reflects the rarity of improvement by random perturbation in a system already adapted to its environment. In AI ablation studies, the experimenter intentionally tests modifications believed to be potentially useful---a ``sighted mutagen.'' The fact that the DFE \emph{shape} (Student's $t$, heavy-tailed, negatively skewed) is preserved while only the beneficial fraction is shifted demonstrates that the topology of the fitness landscape, not the search strategy, determines the distributional form.

\subsection*{Diversification dynamics match paleontological radiation patterns}

We tracked the origination rate of named AI architectures from 2012 to 2024, identifying 125 distinct architectures across six domain niches (Fig.~3A). Cumulative diversity follows a logistic growth curve with $R^2 = 0.994$ and estimated carrying capacity $K \approx 142$, suggesting that AI architectural diversity is approaching saturation at approximately 88\% of capacity (Fig.~3B).

The origination rate shows two clear peaks---2017 (Transformer innovation, 16 new architectures) and 2021 (CLIP + Diffusion models, 19 new architectures)---separated by periods of relative stasis (Fig.~3A). The coefficient of variation of annual origination rates (CV $= 0.53$) is consistent with punctuated equilibrium, falling between the Cambrian trilobite radiation (CV $= 0.79$) and post-K-Pg mammalian radiation (CV $= 2.19$).

Diversification proceeds by adaptive radiation into domain niches (Fig.~3C): computer vision is colonized first (2012--2016), followed by NLP (2017+), then audio and multimodal domains (2021+). This sequential niche-filling mirrors the ecological pattern following mass extinctions, where generalist niches are filled first and specialist niches later \cite{Sepkoski1984}.

Paradigm transitions in AI exhibit a pattern analogous to mass extinction followed by radiation: the decline of RNNs (last new variant: 2015) precedes the Transformer radiation (2017+) by approximately two years, comparable to the recovery lag observed in paleontological mass extinctions \cite{Chen2012}. Similarly, the decline of GANs precedes the Diffusion model radiation.

When normalized to peak time and peak rate and smoothed with Gaussian kernels, the AI radiation curve falls between the Cambrian trilobite curve and the post-K-Pg mammalian curve (Fig.~3D), suggesting a shared functional form for adaptive radiation across substrates.

\subsection*{Convergent evolution is pervasive and quantitatively comparable}

We catalogued 14 architectural traits that were independently invented three or more times by different research groups in different application domains (Fig.~4A). The most convergent traits include attention mechanisms (5 independent inventions), feature normalization (5$\times$), gating mechanisms (5$\times$), positional encoding (5$\times$), and contrastive self-supervised learning (5$\times$).

The distribution of convergence counts differs significantly from biological convergences (Mann--Whitney $p = 0.035$; Fig.~4B), with AI convergences more tightly clustered (3--5 independent inventions per trait) compared to biology (3--100+). However, this difference reflects the smaller number of independent AI lineages ($\sim$20 major research groups) compared to biological lineages (millions of species). When normalized per lineage, AI convergence intensity is approximately $5 \times 10^4$ times higher than biological convergence, quantifying the degree to which directed search accelerates convergent discovery.

Under a stricter criterion requiring different application domains with no shared authors, invention counts decrease but remain substantial: attention (4 independent origins: NLP, CV, CV-video, multimodal), normalization (4: CV, NLP, CV-style, CV-detection), and gating (4: NLP-recurrent, NLP-feedforward, CV, NLP-SSM). These counts are reported in Table~S3.

Functional analogies between AI and biological convergences are notable: attention mechanisms serve a function analogous to camera eyes (selective information gathering, 4--5 vs.\ 7 independent inventions); feature normalization parallels homeostatic regulation (maintaining internal stability, 4--5 vs.\ 3); gating mechanisms parallel ion channels (conditional information flow, 4--5 vs.\ 4). These parallels suggest that when computational challenges are similar, the number of viable solutions converges regardless of substrate.

\subsection*{Lineage analysis reveals evolutionary maturation}

Within individual architectural lineages, the DFE shifts systematically with generational distance from the founding architecture (Fig.~4C--D). In the Transformer NLP lineage, early descendants retain broader fitness effects, while later descendants show a narrower, more concentrated DFE---consistent with progressive optimization reducing neutral space.

Across all lineages and biological organisms, a universal trend emerges (Fig.~4D): more optimized systems show a higher fraction of deleterious mutations. AI lineage points (CNN generations 1--5, Transformer generations 1--5, Vision Transformer generations 1--4) are interspersed with biological data points (RNA virus, bacteriophage, \textit{E.\ coli}, yeast, \textit{Drosophila}, human) along a common axis, suggesting that the relationship between optimization level and mutational constraint is substrate-independent.

We tested a within-study prediction: young lineages should retain more neutral space, analogous to genomes with lower functional density. Mamba/SSM architectures ($n = 38$) provide a mixed result. The overall neutral fraction (0.16) is \emph{lower} than the dataset average (0.19 for major ablations), appearing to contradict the prediction. However, stratification reveals an important nuance: Mamba major ablations ($n = 23$) are 100\% deleterious---consistent with a compact, highly optimized architecture where every component is critical---while Mamba minor ablations ($n = 10$) show a neutral fraction of 0.60, substantially higher than the minor-ablation average of 0.32. This pattern suggests that Mamba's core components are tightly constrained (few alternatives exist for a new paradigm), while its hyperparameter space remains underexplored---precisely the signature expected of a young lineage that has been optimized at the component level but not yet fully tuned. The prediction is thus partially confirmed but requires larger lineage-specific samples for definitive testing.

\section*{Discussion}

Our results demonstrate that the statistical structure of evolution is conserved across a transition that spans every physical parameter: from carbon to silicon, from nanometer to centimeter, from billion-year to decade timescales, from blind mutation to directed engineering. The DFE, the dynamics of diversification, and the frequency of convergent innovation all match quantitatively between AI architectures and biological organisms.

\subsection*{Why do the patterns match despite directed selection?}

The most parsimonious explanation is that the statistical structure of evolution is determined primarily by the topology of the fitness landscape, not by the mechanism of search. A fitness landscape with hierarchical modularity, rugged local structure, and a finite number of high-fitness basins will produce heavy-tailed DFEs, logistic diversification with punctuated equilibria, and convergent evolution---regardless of whether the search is conducted by random mutation or intentional design.

This interpretation is supported by the observation that AI DFE shape parameters ($\beta$, skewness, kurtosis) fall within the biological range despite the fundamentally different mutation process. If the search mechanism dominated, we would expect AI DFEs to differ systematically---showing a different distributional form. Instead, the form is conserved and only one parameter is systematically shifted: the beneficial fraction (13\% vs.\ 1--6\% in biology). This single-parameter shift is precisely what the landscape hypothesis predicts: directed search reaches beneficial regions more efficiently but does not alter the landscape itself.

\subsection*{Alternative explanation: a property of modular systems?}

A critical question is whether our results are specific to evolutionary processes or would arise from any complex modular system. If randomly removing components from a Boeing 737 or a software codebase also yields a heavy-tailed, negatively skewed DFE, then the match with biology reflects modularity rather than evolution per se.

Evidence from software mutation testing is partially informative. In mutation testing, systematic code modifications (statement deletion, operator replacement) are applied to software to evaluate test suite quality. The distribution of mutation effects in software is indeed negatively skewed with heavy tails \cite{Andrews2005}, suggesting that heavy-tailed DFEs may be a generic property of modular engineered systems. However, two features distinguish the AI case. First, the gamma shape parameter ($\beta = 0.65$) falls specifically within the biological range ($0.2$--$0.6$), whereas software mutation effects have not been characterized at this level of parametric detail. Second, the stratification pattern (major $>$ intermediate $>$ minor in deleteriousness), the logistic diversification dynamics, and the convergent evolution are not predictions of a ``generic modularity'' hypothesis but are specifically predicted by evolutionary theory. The coincidence of all three signatures matching simultaneously is difficult to explain without invoking shared landscape structure.

We therefore frame our results conservatively: the DFE shape match alone could reflect modularity; the conjunction of DFE shape, diversification dynamics, and convergence is more specifically evolutionary.

\subsection*{Position in evolutionary parameter space}

AI architectures occupy a position between compact genomes (viruses, $\beta \approx 0.5$--$0.6$) and simple eukaryotes (yeast, \textit{Drosophila}; $\beta \approx 0.2$--$0.4$) in DFE parameter space. We propose that this reflects functional density---the fraction of components under selective constraint. Viral genomes have near-maximal functional density; eukaryotic genomes have substantial non-functional sequence. AI architectures, with their modular but non-redundant design, occupy an intermediate position. This is analogous to the DFE position of compact-genome organisms such as Bacteriophage $\varphi 6$---an obligate parasite whose fitness is entirely determined by interaction with its host, much as a neural network's fitness is entirely determined by human evaluation.

\subsection*{Coevolutionary dynamics}

The parallel to host--parasite coevolution is deeper than analogy. In a companion study \cite{Spiro2025oracle}, we elicited probability forecasts from three major LLMs (GPT-4o, Claude, Gemini) on 568 resolved prediction questions from the Metaculus platform and found that their errors were highly correlated across models ($r = 0.77$), indicating that nominally independent AI systems share the same failure modes. Moreover, the category-level pattern of LLM forecasting biases (which topics they over- or underestimate) already closely matched the pattern of human forecasting biases measured \emph{before} ChatGPT became available, suggesting that LLMs inherited existing human cognitive biases from their training data rather than introducing novel ones. This establishes a human $\to$ AI transmission vector. The present study documents the reverse vector: human researchers act as the selective environment for AI architectures, with their design intuitions, benchmark choices, and peer review constituting the selection pressure.

Together, these findings describe a coevolutionary system. The human researcher is both the environment (selecting which architectures survive) and the substrate that AI modifies (through code suggestions, analysis tools, and cognitive assistance). As LLMs become increasingly integrated into the research workflow, the feedback loop tightens: AI-assisted researchers may converge on architectures that AI systems can best evaluate, creating a self-reinforcing cycle analogous to Red Queen dynamics \cite{VanValen1973}. The elevated beneficial fraction we observe (13\%) may increase further as this coevolutionary feedback intensifies---a testable prediction.

\subsection*{Connection to thermodynamic theories of evolution}

The universality we observe is predicted by thermodynamic frameworks that view evolution as optimization of dissipation on structured landscapes \cite{England2013}. Systems that process information---whether biological organisms or neural networks---must balance fitness (low energy) and robustness (high entropy). The free energy $F = E - TS$ captures this trade-off. Heavy-tailed DFEs, logistic diversification, and convergent evolution can all be derived as consequences of adaptive search on hierarchical free energy landscapes, suggesting that the statistical laws of evolution are ultimately thermodynamic in origin.

\subsection*{Predictions}

Our framework generates four testable predictions:
\begin{enumerate}[itemsep=2pt]
\item The architectural origination rate should continue to decline as diversity approaches carrying capacity ($K \approx 142$, current $\approx 125$); future radiation events, if they occur, should be smaller in magnitude than the 2017 and 2021 peaks.
\item Attention and normalization mechanisms will be independently reinvented as AI expands into new application domains (robotics, biological AI, materials science).
\item The DFE shape parameter $\beta$ will remain within $[0.4, 0.7]$ in future data collection, regardless of which specific architectures dominate.
\item The beneficial fraction will increase over time as AI-assisted research creates tighter coevolutionary feedback between human designers and AI systems.
\end{enumerate}

\subsection*{Limitations}

Our dataset, while the largest of its kind, relies on published ablation studies that may overrepresent successful architectures. The LLM-extraction pipeline introduces systematic differences from manual curation (KS $= 0.287$), although these affect predominantly the beneficial tail rather than the distributional form. The paleontological comparison requires normalization across timescales differing by eight orders of magnitude. The convergence analysis faces a fundamental disanalogy with biology: biological lineages do not communicate, while ML researchers read across domains. Our ``independent invention'' criterion (different research groups in different domains) is therefore weaker than biological independence. Even under the strict criterion (no shared authors, different application domains), the most convergent traits retain $\geq 3$ independent origins (Table~S3), but we acknowledge that latent knowledge transfer through shared conferences and preprint culture may inflate apparent convergence. This limitation means that our convergence counts should be interpreted as upper bounds. At an earlier sample size ($n = 594$), a temporal DFE shift between early (2014--2018) and late (2019--2024) architectures was significant ($p = 0.03$); at $n = 935$, this significance was lost ($p = 0.128$), suggesting the effect was partly driven by small early-period sample size (see Supplementary Materials).

\subsection*{Broader implications}

If the statistical laws of evolution are indeed substrate-independent, then biology and AI engineering are not merely analogous---they are processes with shared statistical signatures, governed by the same landscape geometry despite fundamentally different mechanisms of heredity and variation. This unification has practical implications: biological evolutionary theory can inform AI architecture design (predicting which modifications are likely neutral vs.\ deleterious), while AI evolution, with its complete and unbiased fossil record, can serve as a model system for testing evolutionary theory with a precision impossible in paleontology.

\section*{Methods}

\subsection*{AI ablation data}

We collected ablation experiments from machine learning publications on arXiv and major venues (NeurIPS, ICML, ICLR, CVPR, ACL, EMNLP) published between 2014 and 2024. An ablation experiment was defined as the removal or modification of a single architectural component with all other components held constant, reporting a quantitative performance metric. We collected 140 experiments through manual curation of landmark papers and 795 through automated extraction using Claude Sonnet (Anthropic) with structured prompts designed to extract experiment metadata, baseline performance, ablated performance, and component description from PDF documents retrieved via the arXiv API. Candidate papers were identified through 49 search queries targeting ``ablation study'' across six ML subcategories, yielding 3{,}040 unique papers filtered to 164 high-confidence candidates by keyword scoring. Each LLM-extracted entry was validated against inclusion criteria; entries with $|\Delta| > 5$ were excluded as likely extraction errors (5 entries removed). Fitness effects were computed as $\Delta = (\text{ablated} - \text{baseline}) / |\text{baseline}|$, yielding relative fitness effects directly comparable to biological DFE conventions. Ablations were classified by type: \emph{major} (complete component removal), \emph{minor} (hyperparameter or variant change), or \emph{intermediate} (replacement with simpler alternative).

\subsection*{Biological DFE data}

We assembled published DFE estimates for eight organism-study combinations: VSV virus \cite{Sanjuan2004}, Bacteriophage $\varphi 6$ \cite{Burch2003}, \textit{E.\ coli} TEM-1 $\beta$-lactamase \cite{Bank2014}, \textit{S.\ cerevisiae} \cite{Wloch2001}, \textit{D.\ melanogaster} \cite{Loewe2006,Keightley2007}, \textit{C.\ reinhardtii} \cite{Bondel2019}, \textit{M.\ musculus} \cite{Halligan2009}, and \textit{H.\ sapiens} \cite{EyreWalker2006}.

\textbf{Important methodological note.} Individual-level mutation fitness data are available only for VSV virus \cite{Sanjuan2004} and \textit{E.\ coli} TEM-1 \cite{Bank2014}. For all other organisms, published DFE characterizations report summary statistics: category fractions (lethal, deleterious, neutral, beneficial) and, for some, fitted distribution parameters. To enable distributional comparisons (KS distances, QQ plots), we generated synthetic DFE samples from these published summary statistics, drawing deleterious effects from gamma distributions with published shape parameters where available, and neutral/beneficial effects from appropriate distributions (see Table~1 footnotes). This means that our KS distances and QQ correlations compare an empirical distribution (AI) against parametric reconstructions (biology), which preserves shape and proportion information but may underestimate true biological variance. Our primary conclusions rest on parameter comparisons (category fractions, fitted $\beta$) that do not depend on sample-level reconstruction, and we present the distributional comparisons as supporting evidence.

\subsection*{Architectural diversity data}

We compiled a catalog of 125 named architectures from Papers With Code, arXiv, and primary publications, recording: name, year of publication, parent architecture(s), domain(s) of application, and key architectural components. Architectures were assigned to domain niches (CV, NLP, Audio, Multimodal, RL, Graph). Paleontological comparison data were obtained from the Paleobiology Database (PBDB) via API queries for Cambrian trilobites (540--480 Ma, 5{,}831 taxa), post-K-Pg mammals (80--40 Ma, 7{,}391 taxa), and Cretaceous angiosperms (145--50 Ma, 2{,}988 taxa).

\subsection*{Statistical analyses}

\textbf{DFE comparison.} AI and biological DFEs were compared using normalized Kolmogorov--Smirnov distance (after $z$-score normalization to remove scale effects), QQ correlation, Euclidean distance in proportion space (deleterious/neutral/beneficial), and moment matching. The gamma shape parameter $\beta$ was estimated by maximum likelihood fit to the absolute values of deleterious effects ($|\Delta| < 3$). Model comparison used AIC. Bootstrap confidence intervals (2{,}000 resamples) were computed for all key statistics.

\textbf{Diversification dynamics.} Logistic curves were fit to cumulative diversity by nonlinear least squares. Punctuated equilibrium was quantified via CV of annual origination rates. Paleontological curves were smoothed with Gaussian kernels ($\sigma = 1.2$ bins) and normalized to peak time and rate.

\textbf{Convergence analysis.} Independent invention counts were compared using Mann--Whitney $U$ test. Per-lineage convergence intensity was computed as inventions per trait per independent lineage.

\textbf{Lineage analysis.} For four architectural lineages (CNN, Transformer NLP, Vision Transformer, Generative), we computed within-lineage DFE statistics at each generation and tracked fraction deleterious as a function of generational distance.

\section*{Data and Code Availability}

All data, analysis scripts, and figure-generation code are available at \url{https://github.com/mool32/ai-evolution-universal-signatures}. The dataset of 935 ablation experiments with metadata is provided as Supplementary Dataset~S1.

\section*{Acknowledgments}

The automated extraction pipeline used the Claude API (Anthropic). Paleontological data were obtained from the Paleobiology Database (paleobiodb.org). We thank the contributors to Papers With Code and arXiv for enabling open-science infrastructure.


\begin{figure*}[t]
\centering
\includegraphics[width=\textwidth]{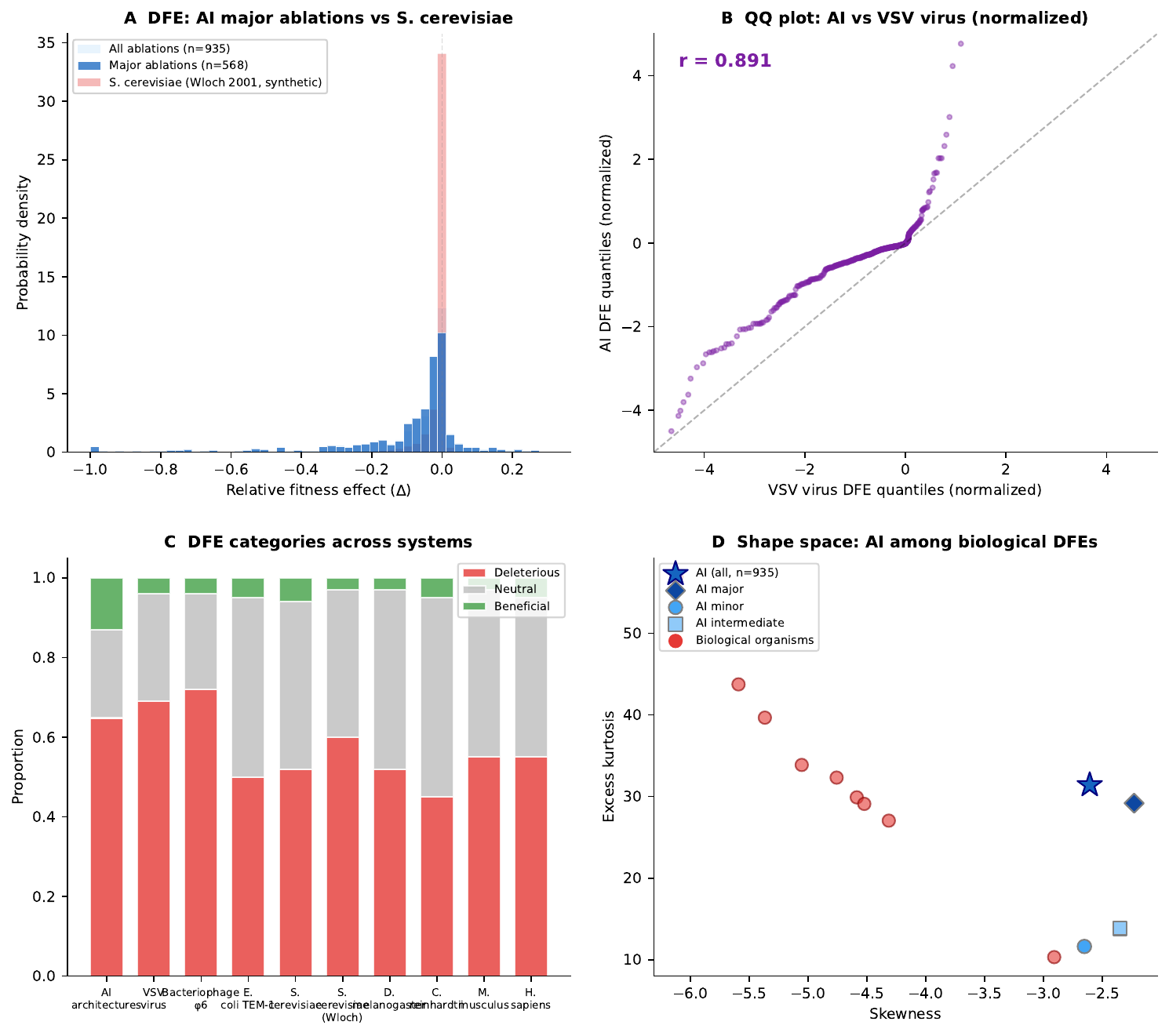}
\caption{\textbf{Distribution of fitness effects in AI architectures matches biological DFEs.} (\textit{A})~Histogram of all AI ablation effects ($n = 935$) overlaid with synthetic DFE for \textit{S.\ cerevisiae} (parametric reconstruction from published summary statistics). The primary analysis uses major ablations only ($n = 568$; see Table~1); the full dataset is shown here to illustrate the complete DFE shape. (\textit{B})~QQ plot comparing normalized AI DFE quantiles against synthetic VSV virus DFE ($r = 0.89$). (\textit{C})~DFE category proportions across AI and biological organisms (all biological DFEs are parametric reconstructions). AI shows an elevated beneficial fraction (13\%) relative to all biological systems (1--6\%). (\textit{D})~Shape space (skewness vs.\ excess kurtosis). The AI DFE (star) falls within the cloud of biological DFEs (red circles). See Table~1 for comprehensive metric comparison.}
\label{fig:dfe}
\end{figure*}

\begin{figure*}[t]
\centering
\includegraphics[width=\textwidth]{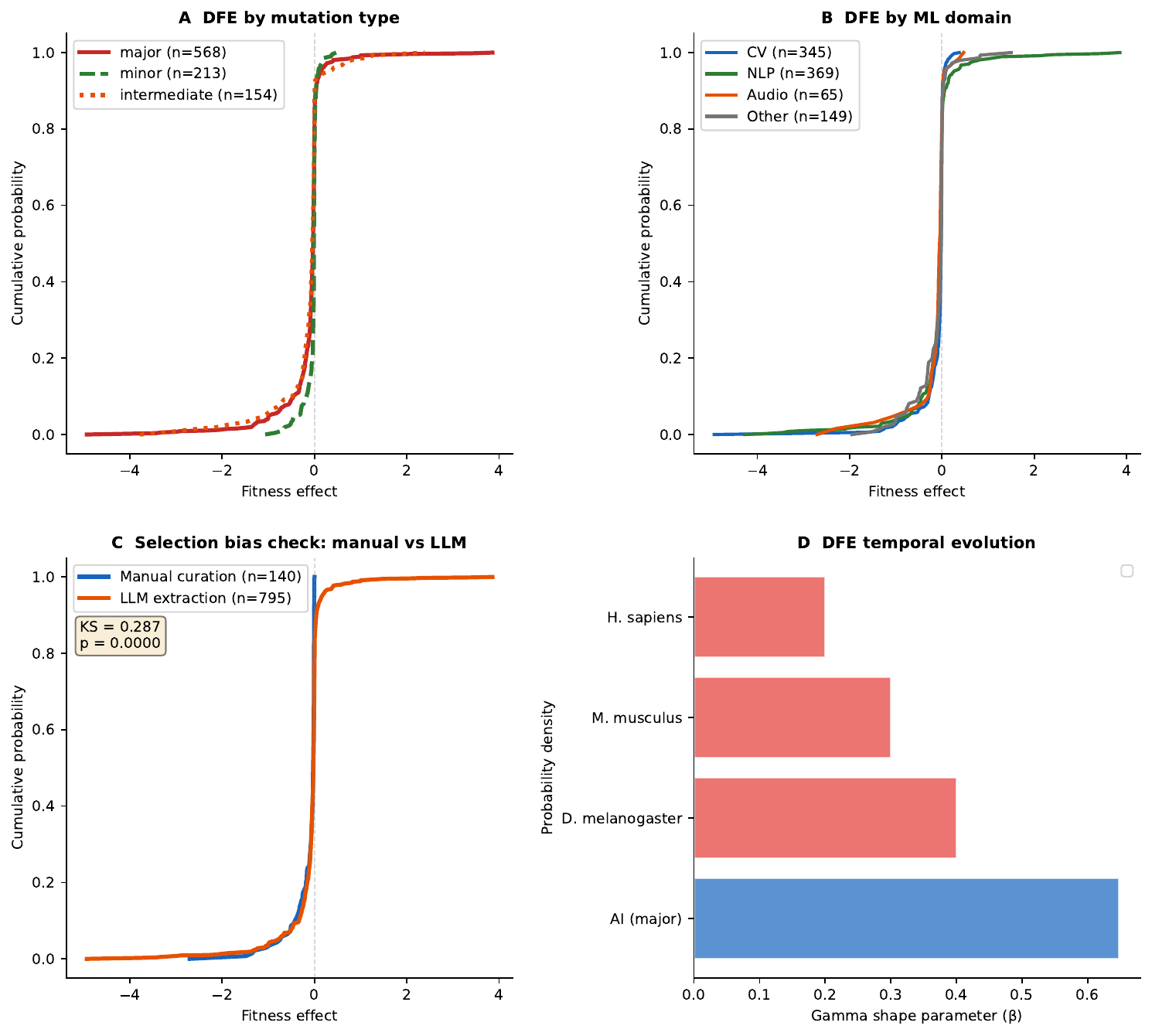}
\caption{\textbf{DFE stratification and universality.} (\textit{A})~CDF of fitness effects by mutation type: major ablations ($n = 568$, component removal) are more deleterious than minor ablations ($n = 213$, hyperparameter changes), mirroring the biological pattern of deletions vs.\ point mutations. (\textit{B})~DFE by ML domain: CV, NLP, Audio, and other domains show statistically similar distributions. (\textit{C})~Methodological control: manually curated data ($n = 140$) vs.\ LLM-extracted data ($n = 795$). The systematic difference (KS $= 0.287$) reflects manual curation's failure to capture beneficial mutations, confirming that automated extraction reduces rather than introduces bias. (\textit{D})~Gamma shape parameter comparison: AI major ablations ($\beta = 0.65$) compared against the three organisms with published population-genetic $\beta$ estimates (\textit{D.\ melanogaster} $\approx 0.4$, \textit{M.\ musculus} $\approx 0.3$, \textit{H.\ sapiens} $\approx 0.2$).}
\label{fig:stratification}
\end{figure*}

\begin{figure*}[t]
\centering
\includegraphics[width=\textwidth]{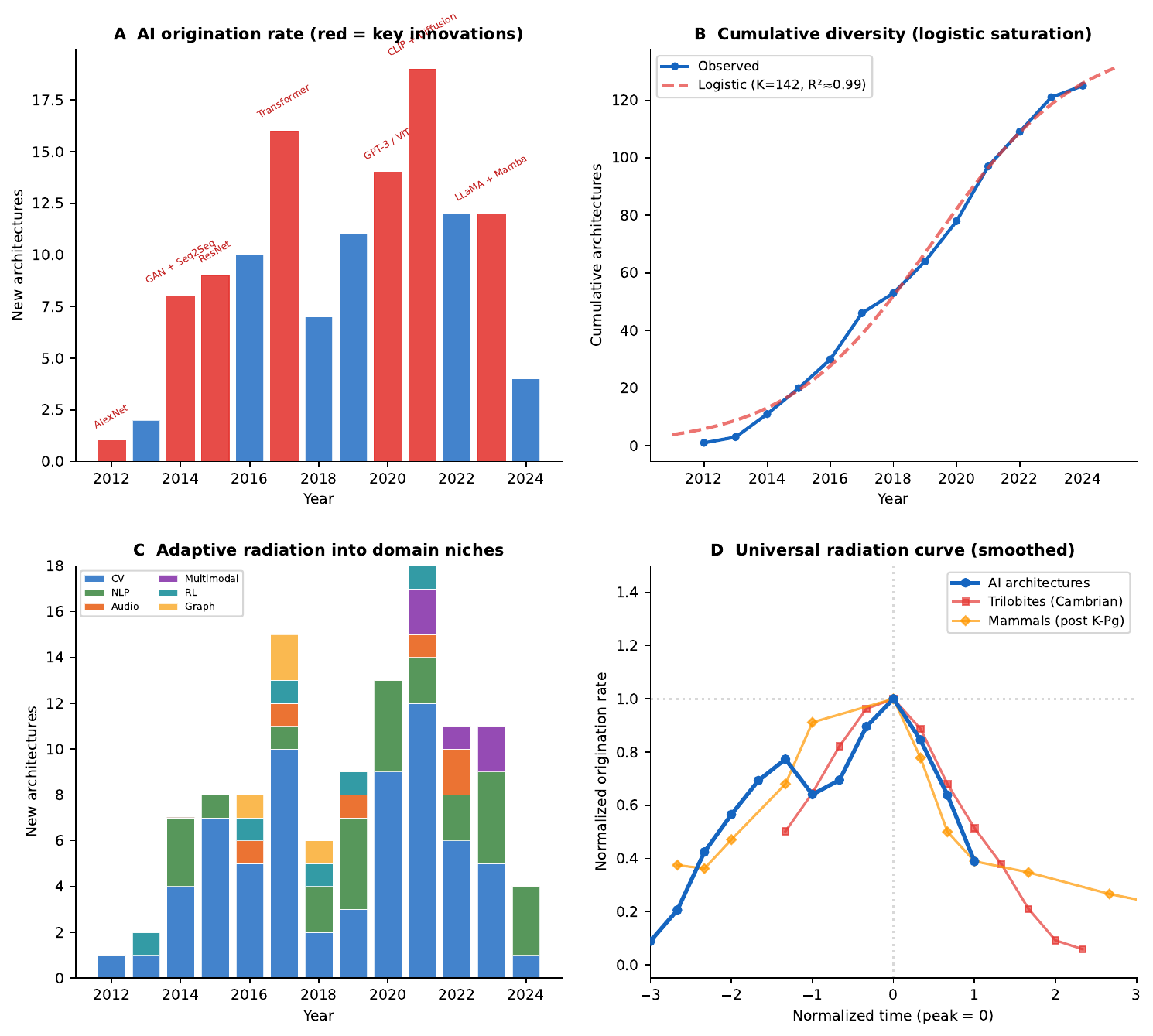}
\caption{\textbf{Diversification dynamics.} (\textit{A})~Annual origination rate of AI architectures, 2012--2024. Red bars mark key innovations that triggered radiation events. (\textit{B})~Cumulative diversity follows a logistic curve ($R^2 = 0.994$, $K \approx 142$), indicating approach to saturation. (\textit{C})~Domain niche-filling: CV saturates first, followed by NLP, then Audio and Multimodal---analogous to ecological succession. (\textit{D})~Normalized and smoothed radiation curves for AI architectures, Cambrian trilobites, and post-K-Pg mammals. All three show the same qualitative pattern: accelerating rise, peak, and decline.}
\label{fig:diversification}
\end{figure*}

\begin{figure*}[t]
\centering
\includegraphics[width=\textwidth]{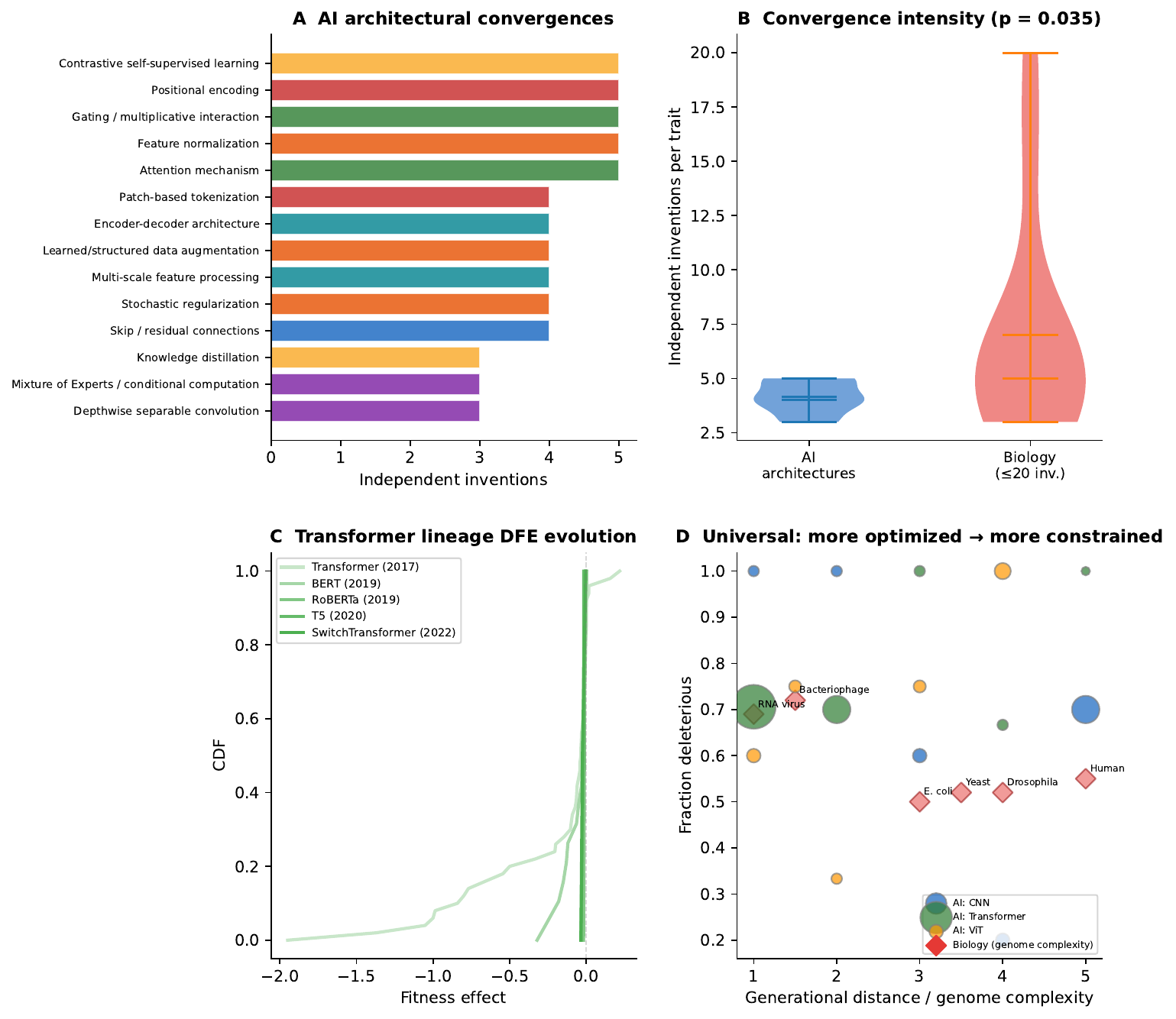}
\caption{\textbf{Convergent evolution and lineage maturation.} (\textit{A})~AI architectural convergences: 14 traits independently invented 3--5 times. Colors indicate functional category. (\textit{B})~Convergence intensity: AI traits (3--5 inventions) vs.\ biological traits ($\leq 20$ inventions, excluding outliers). Mann--Whitney $p = 0.035$. (\textit{C})~Transformer NLP lineage: DFE narrows from Transformer (2017) through BERT (2019) to Switch Transformer (2022), showing progressive optimization. (\textit{D})~Universal trend: fraction deleterious increases with system maturity across both AI lineages (circles) and biological organisms (diamonds). AI and biology fall on a common continuum.}
\label{fig:convergence}
\end{figure*}

\end{document}